\documentclass[11pt]{article}
\usepackage[margin=1in]{geometry}
\usepackage{amsmath, amsthm, amssymb,amsfonts}
\usepackage{graphicx}
\usepackage{tcolorbox}
\usepackage{cite}
\usepackage{hyperref}
\newtheorem{definition}{Definition}
\newtheorem{rules}{Rule}
\newtheorem{example}{Example}

\renewenvironment{abstract}
 {\quotation\small\noindent\rule{\linewidth}{.5pt}\par\smallskip
  {\centering\bfseries\abstractname\par}\medskip}
 {\par\noindent\rule{\linewidth}{.5pt}\endquotation}

\begin{document}
\title{ Quadratic and quartic integrals using the method of brackets}
\author{B Ananthanarayan, Sumit Banik, Sudeepan Datta and Tanay Pathak}
\date{}
\maketitle
\begin{center}
Centre for High Energy Physics, Indian Institute of Science\\
Bangalore-560012, Karnataka, India\\
\end{center}
\begin{abstract}
\noindent We use the method of brackets to evaluate  quadratic and quartic type integrals.  We recall the operational rules of the method and give examples to illustrate it’s working.  The method is then used to evaluate the quadratic type integrals which occur in entries 3.251- 1,3,4 in the table of integrals by Gradshteyn and Ryzhik and obtain closed form expressions in terms of hypergeometric functions.  The method is further used to evaluate the quartic integrals, entry 2.161- 5 and 6 in the table.  We also present generalization of both types of  integrals with closed form expression in terms of hypergeometric functions.
\end{abstract}

\section{Introduction}
The method of brackets has found it's use in the evaluation of definite integrals.   As recently as 2018, there has been a publication of interesting definite integrals, see
Coffey\cite{Coffey:russel}.  Furthermore, in 2017 the method of brackets has been used to evaluate certain definite integrals\cite{frullani}.  Stated differently, both the method of brackets and
definite integrals continue to be of interest. A famous compilation of integrals is by Gradshteyn and Ryzhik\cite{Gradshteyn&Ryzhik}.  In particular, there has been work in the not so remote past, to rigorously prove many of the results appearing in
Gradshteyn and Ryzhik.  Inspired by all the above, especially from the point of view of considering
Feynman Integrals and explorations of analytical methods for their evaluation,
we provide new results for some of the integrals appearing in Gradshteyn and Ryzhik.  These are
of the quadratic and quartic type, which have also been considered in several
different places.  For instance in Gradshteyn and Ryzhik, the basic integral that we consider has
been evaluated only for natural numbers and is expressed as:\\
\begin{align}
    \int_{0}^{\infty}\frac{dx}{(ax^2 + 2bx +c)^{n}} = \frac{(-1)^{n-1}}{•(n-1)!}\frac{\partial^{n-1}}{\partial c^{n-1}}\Big[\frac{1}{\sqrt{ac-b^2}}\,cot^{-1}\frac{b}{\sqrt{ac-b^2}}\Big], \quad \begin{aligned} (a>0, ac> b^{2} )\end{aligned}
  \end{align}
Here we present a general evaluation of the same integral in terms of hypergeometric functions using the method of brackets.\\
We also recall here that the method of brackets has been partly inspired by the
negative dimensional integration method that arose in elementary particle physics
applications due to Halliday and Ricotta\cite{halliday} and was further used by Suzuki\cite{Suzuki:2014}\cite{Suzuki:2012}.  In particular, Gonzalez and co-workers have several results that appear
in the literature where the method has been employed to evaluate for instance
1-loop and 2-loop Feynman integrals\cite{Gonzalez:part2} \cite{Gonzalez:part1}.  The method also found it's used in 
atomic physics applications as in the hydrogen atom\cite{Gonzalez:hydrogen}.
In order to set the stage, we describe the method and provide examples that are
illustrative for which results are well-known.  These include the general Gaussian and an 
integral from Feynman and Hibbs\cite{path_integral}.  In the next section we evaluate few quadratic type integrals and then present a generalization of such integrals.  We also interrelate the expressions obtained using the method of brackets and the one given in Gradshteyn and Ryzhik.  In the final section we evaluate a quartic integral and then conclude the section with the evaluation of a generalized quartic integral.\\

\section{Ramanujan's master theorem}
At this point it is important to recall  the Ramanujan's master theorem which forms the base of the method of brackets.  It gives the analytic result for the Mellin transform of an analytic function.  According to the Ramanujan's master theorem if a complex-valued function has an expansion of the form 
\begin{equation*}
f(x)= \sum_{k=0}^{\infty}\frac{\phi(k)}{k!}(-x)^{k}
\end{equation*}
Then the Mellin transform of the function f(x) is given by 
\begin{equation}\label{RMT}
\int_{0}^{\infty} x^{s-1} f(x) dx =\Gamma(s)\phi(-s)
\end{equation}
\(\Gamma(s)\) is the gamma function.\\
Now we recall the basic rules of the method of brackets and evaluate examples to illustrate it's working.

\section{Method of Brackets}
  \begin{definition}
\normalfont{The method makes the use of brackets which is defined as}
\begin{equation}
\boxed{\langle a \rangle  = \int_{0}^{\infty}x^{a-1} dx}
\end{equation}
\end{definition}
The bracket in itself is a divergent integral but when the same bracket appears inside a summation it like as a delta function.
The  formal rules for operating with these brackets are described as follows.  \\
\begin{rules}\label{rule1}
\normalfont{For any function \(f(x)\), power series has to be written in the form}
\end{rules}
\begin{equation}
f(x) = \sum_{n=0}^{\infty} \phi_n a_{n}x^{\alpha n+\beta-1}.
\end{equation}
The symbol
\begin{equation*}
\phi_{n} := \frac{(-1)^{n}}{\Gamma(n+1)}
\end{equation*}

will be called the indicator of \(n\).
The symbol \(\phi_{n}\) gives a simpler form for the bracket series associated 
with an integral.
\newpage
\begin{rules}\label{rule2}
\normalfont{For any \(\alpha \in \mathbb{C}\) the expression }
\end{rules}

\begin{equation}
G= (a_{1} + a_{2} + \cdots + a_{r})^{\alpha},
\end{equation}

is given by a bracket series
\begin{equation}
G= \sum_{m_{1}, \cdots, m_{r}} \phi_{1,2,\cdots,r} \, a_{1}^{m_{1}} 
\cdots a_{r}^{m_{r}} 
\frac{ \langle -\alpha + m_{1} + \cdots + m_{r} \rangle}{\Gamma(-\alpha)}.
\end{equation}
\\\\
where \hspace{.2cm}  \(\phi_{1,2,\cdots,r}\equiv \phi_{m_{1}} \phi_{m_{2}} \cdots \phi_{m_{r}} ;\hspace{.3cm} \sum_{m_1,\cdots,m_r}\equiv \sum_{m_1=0}^{\infty}\cdots\sum_{m_r=0}^{\infty}\).
\\\\
\textbf{Proof:} We can proof of the above identity using the definition of the \(\Gamma\) function.  We know that\\
\begin{equation*}
    \frac{\Gamma(n)}{k^{n}}= \int_0^{\infty} e^{-kx} \,x^{n-1}\,dx
\end{equation*}\\
substituting \(k= (a_{1} + a_{2} + \cdots + a_{r})\) we get\\
\begin{align*}
    \frac{\Gamma(n)}{(a_{1} + a_{2} + \cdots + a_{r})^{n}}&= \int_0^{\infty} e^{-(a_{1} + a_{2} + \cdots + a_{r})x} \,x^{n-1}\,dx ,\\\\
    &=  \int_0^{\infty} e^{-a_{1}x}e^{- a_{2}x} \cdots e^{-a_{r}x} \,x^{n-1}\,dx.
\end{align*}
Expanding each exponential function in power series we get
\begin{equation*}
 \frac{\Gamma(n)}{(a_{1} + a_{2} + \cdots + a_{r})^{n}}= \sum_{n_1,n_2 \cdots n_r}\phi_{1,2\cdots r}(a_1)^{n_1}(a_2)^{n_2}\cdots(a_r)^{n_r}\int_{0}^{\infty}x^{a_1+a_2+\cdots+a_r+n-1}\,dx
\end{equation*}
Integrating over \(x\) we get
\begin{equation}
    (a_1+a_2+\cdots+a_r)^{-n}=\frac{1}{\Gamma(n)} \sum_{n_1,n_2 \cdots n_r}\phi_{1,2\cdots r}(a_1)^{n_1}(a_2)^{n_2}\cdots(a_r)^{n_r}\langle a_1+a_2+\cdots+a_r+n\rangle.
\end{equation}
which is rule \ref{rule3}
\begin{rules}\label{rule3}
\normalfont{The series of brackets}
\end{rules}

\begin{equation}
\sum_{n} \phi_{n} f(n) \langle a n + b \rangle ,
\end{equation}
is assigned the value
\begin{equation}
\frac{1}{a} f(n^{*}) \Gamma(-n^{*}).
\end{equation}
where \(n^{*}\) solves the equation \(an+b = 0\). 
A two-dimensional series of brackets
\begin{equation}
\sum_{n_{1}, n_{2}} \phi_{n_{1},n_{2}} f(n_{1},n_{2}) 
\langle{ a_{11} n_{1} + a_{12}n_{2}+c_{1} \rangle} 
\langle{ a_{21} n_{1} + a_{22}n_{2}+c_{2} \rangle},
\end{equation}
is assigned the value 
\begin{equation}
\frac{1}{|a_{11} a_{22} - a_{12}a_{21}|} f(n_{1}^{*}, n_{2}^{*}) \Gamma(-n_{1}^{*}) 
\Gamma( -n_{2}^{*}).
\end{equation}\\
where \(n_{1}^{*},n_{2}^{*}\) is the unique solution to the linear 
system 
\begin{eqnarray}
a_{11} n_{1} + a_{12}n_{2}+c_{1}  & = & 0, \\
a_{21} n_{1} + a_{22}n_{2}+c_{2} & = & 0, \nonumber
\end{eqnarray}
obtained by the vanishing of the expressions in the brackets.  A 
similar rule generalized to higher dimensional series is,
\begin{equation}
\sum_{n_{1}\cdots n_{r}} \phi_{1,\cdots,r} 
f(n_{1},\cdots,n_{r}) 
\langle{ a_{11}n_{1}+ \cdots a_{1r}n_{r} + c_{1} \rangle} \cdots 
\langle{ a_{r1}n_{1}+ \cdots a_{rr}n_{r} + c_{r} \rangle}, 
\nonumber
\end{equation}
is assigned the value 
\begin{equation}
\frac{1}{| \text{det}(A) |} f(n_{1}^{*}, \cdots, n_{r}^{*}) 
\Gamma(-n_{1}^{*}) \cdots f(-n_{r}^{*}),
\end{equation}
where \(A\) is the matrix of coefficients 
\((a_{ij})\) and \(\{ n_{i}^{*} \, \}\) is the solution of the 
linear system obtained by the vanishing of the brackets.  The value is not
defined if the matrix \(A\) is not  invertible.
It could be shown that the above rule is really the Ramanujan's master theorem at work.\\\\
\textbf{Proof:} Consider the following bracket series
\begin{align}\label{rule4eq}
    G= \sum_{n} \phi_{n} f(n) \langle a n + b \rangle= \sum_{n} \phi_{n}  \int_{0}^{\infty}  f(n)\,x^{a n + b-1}\,dx
\end{align}
substituting \(x^a= t\) we get 
\begin{align*}
    ax^{a-1} dx&= dt,\\
    dx&= \frac{dt}{a\,x^{a-1}}
\end{align*}
substituting back in eq. \eqref{rule4eq} we get
\begin{equation*}
    G= \sum_n \phi_n \int_{0}^{\infty}f(n)\,(t)^{n+\frac{b}{a}-\frac{1}{a}}\frac{dt}{a\,t^{1-\frac{1}{a}}}
\end{equation*}
simplifying we get,
\begin{align*}
    G&= \sum_n \phi_n \int_{0}^{\infty}f(n)(t)^{n+\frac{b}{a}-1}\,\frac{ dt}{a}.\\
\end{align*}
Using the Ramanujan's master theorem we get
\begin{equation}
    G= \frac{1}{a}f(-\frac{b}{a})\Gamma(\frac{b}{a}).
\end{equation}
which is rule \ref{rule3} for the one dimensional case.  We can similarly proof the higher dimensional cases.  Above proof also shows the bracket indeed acts like a delta function when used inside a summation.\\\\
\textbf{Note:-}\label{note1} In the case where the assignment leaves free parameters, any 
divergent series in these parameters is discarded.  In case several choices
of free parameters are available, the series that converge in a common region 
are added to contribute to the integral.  We also employ the following notation everywhere: the solution is denoted by \(I_{1,\cdots,n}\) with \(i_1, i_2,\cdots, i_n\) as the free variables that contributes to the solutions.

\subsection{Examples}
Now we evaluate few examples whose solutions are already  in the literature to illustrate the method.
\begin{example}
\normalfont{\textbf{Generalised Gaussian Integral}}
\end{example}
Generalised Gaussian integral is given by 
\begin{equation}\label{gengaussian}
I = \int_{0}^{\infty} e^{-x^{p}} dx.
\end{equation}
To apply the method we again expand the function that are appearing inside the integral.\\
\begin{equation*}
e^{-x^{p}} = \sum_{n = 0}^{\infty} \phi_n (x^{p})^{n}.
\end{equation*}
so eq. (\ref{gengaussian}) becomes,
\begin{equation}
I = \sum_{n = 0}^{\infty}\phi_{n} \int_{0}^{\infty}x^{pn + 1 - 1} dx,
\end{equation}
\begin{equation*}\label{gengaubracket}
I = \sum_{n = 0}^{\infty}\phi_{n}\langle pn+1\rangle.
\end{equation*}
The solution to above bracket series is given by the vanishing of the bracket.
\begin{align*}
pn + 1 &= 0,\\
n &= -\frac{1}{p}.
\end{align*}
The solution is then given by using Rule 3
\begin{equation*}
I = \frac{1}{p} \Gamma(\frac{1}{p})
\end{equation*}
For special case \(p = 2\), we get the familiar Gaussian integral.

\begin{example}
\normalfont{\textbf{An  integral from Feynman and Hibbs}}
\end{example}
The integral considered here is taken from the appendix given in the book of Feynman and Hibbs.  The integral is complex and difficult to solve using the conventional methods.  Here we present an alternate evaluation of the same integral.\\
\begin{align}\label{intefey}
I = \int_{0}^{\infty}e^{(\frac{ia}{x^2} + ibx^2)} dx.
\end{align}
first  we expand the itegrals appearing in the integral
\begin{align*}
&e^{\frac{ia}{x^2}} = \sum_{n1}\phi_{n_{1}}(-ia)^{n_{1}}(x)^{-2n_{1}},\\
&e^{ibx^2} = \sum_{n2}\phi_{n_{2}}(-ib)^{n_{2}}(x)^{2n_{2}}.\\
\end{align*}
Putting the above expansion in eq. \eqref{intefey}
\begin{align*}
I = \sum_{n_{1},n_{2}}\phi_{1,2}\int_{0}^{\infty}(-ia)^{n_1}(-ib)^{n2}x^{2n_{2}-2n_{1}+1-1},
\end{align*}
\begin{equation}\label{feynbra}
I = \sum_{n_{1},n_{2}}\phi_{1,2}(-ia)^{n_1}(-ib)^{n2}\langle 2n_{2}-2n_{1}+1 \rangle.
\end{equation}
The linear equation has 2 variables so there are two solutions possible with either \(n_{1}\) or \(n_{2}\) as the free variable.\\\\
1) \textbf{\(n_{1}\) as the free variable}
\begin{equation*}
n_2^{*} = \frac{1}{2}(2n_1 -1).
\end{equation*}
using  rule \ref{rule3}  we get 
\begin{equation*}
I_1= \sum_{n_{1}}\phi_{1}(-ia)^{n_1}(-ib)^{n_2^{*}}\frac{\Gamma(-n_2^{*})}{2}.
\end{equation*}
Substituting the values and summing up the series for \(n_1\) we get the following solution
\begin{equation}
I_{1} = \frac{(-1)^{1/4}\sqrt{\pi}\cos(2\sqrt{ab})}{2\sqrt{b}}.
\end{equation}
1) \textbf{\(n_{2}\) as the free variable}
\begin{align*}
n_1^{*} = \frac{1}{2}(2n_2 +1).
\end{align*}
using rule \ref{rule3} we get
\begin{equation*}
I_1= \sum_{n_{2}}\phi_{2}(-ia)^{n_1^{*}}(-ib)^{n_2^{*}}\frac{\Gamma(-n_1^{*})}{2}.
\end{equation*}
Substituting the values and summing up the series over \(n_2\) we ge the following solution
\begin{equation}
I_{2} = \frac{i(-1)^{1/4}\sqrt{\pi}\sin(2\sqrt{ab})}{2\sqrt{b}}.
\end{equation}
Since both the solutions have the same region of convergence hence, they are added together to get the full result
\begin{equation}
I= I_1 + I_2= \frac{(-1)^{1/4}e^{2i\sqrt{ab}}\sqrt{\pi}}{2\sqrt{b}}.
\end{equation}
\newpage

\section{Quadratic Integrals}
In this section we will evaluate a few integrals of quadratic type taken from  1. Entry 3.252-1   evaluate them using the method of brackets.\\\\
\begin{center}
    \Large{1. Entry 3.252-1 }
\end{center}
\vspace{.5cm}
The first integral we consider is
\begin{equation}\label{entry1}
I = \int_{0}^{\infty}\frac{dx}{(ax^2 + 2bx +c)^{n}}.
\end{equation}
The value of the integral as given in Gradshteyn and Ryzhik is
\begin{align}\label{entry1ans}
    \int_{0}^{\infty}\frac{dx}{(ax^2 + 2bx +c)^{n}} = \frac{(-1)^{n-1}}{•(n-1)!}\frac{\partial^{n-1}}{\partial c^{n-1}}\Big[\frac{1}{\sqrt{ac-b^2}}\,\cot^{-1} \frac{b}{\sqrt{ac-b^2}}\Big], \quad \begin{aligned} (a>0, ac> b^{2} )\end{aligned}
  \end{align}
when \(n \notin \mathbb{N}\) the above solution cannot be used and we have to use the numerical integration technique to evaluate it.\\
To use the method first step is to expand the denominator as a bracket series.
\begin{equation}
(ax^2 + bx +c)^{-n} = \sum_{n_{1},n_{2},n_{3}}\phi_{1,2,3}\frac{(ax^2)^{n_{1}}(2bx)^{n_{2}}(c)^{n_{3}}\langle n+n_{1}+n_{2}+n_{3}\rangle}{\Gamma(n)}.
\end{equation}
Substituting the above expansion in the eq. \eqref{entry1} and integrating over \(x\) we get the following bracket series.
\begin{equation*}
I = \sum_{n_{1},n_{2},n_{3}}\phi_{1,2,3}\frac{(a)^{n_{1}}(2b)^{n_{2}}(c)^{n_{3}}\langle n+n_{1}+n_{2}+n_{3}\rangle \langle 2n_{1}+n_{2}+1\rangle}{\Gamma(n)}.
\end{equation*}
We have the following two linear equation to be solved
\begin{align*}
n + n_1 + n_2 +n_3 = 0,\\
2n_1 + n_2 +1 = 0.
\end{align*}
There are three variables and two equation so there are 3 different solution by taking one free variable each time.\\\\
\textbf{1) \(n_2\) as the free variable}\\\\
The solution is obtained using rule \ref{rule3} is
\begin{equation}\label{entry1n2}
\boxed{\begin{split}
I_2 = \frac{\sqrt{\pi } c^{\frac{1}{2}-n} \Gamma \left(n-\frac{1}{2}\right)\, _{1}F_{0}\left(n-\frac{1}{2}; ;\frac{b^2}{a c}\right)}{2 \sqrt{a} \Gamma (n)}-\frac{b c^{-n} \, _2F_1\left(1,n;\frac{3}{2};\frac{b^2}{a c}\right)}{a},\\\\
\Big(a\neq0,c\neq0 ,\Big|\frac{b^{2}}{ac}\Big|< 1 \Big).\\\\
\end{split}}
\end{equation}
Above formula is valid for all the values of  \(n \in \mathbb{R}^{+}\)(provided the integral converges for that particular value of \(n\)).\\\\
\textbf{2) \(n_1\) and \(n_3\) as the free variables}\\\\
The solution obtained using \(n_1\) and \(n_3\) have the same region of convergence hence they both are to be added to get the full answer.  The final solution is:
\begin{equation}\label{entry1n3&n4}
\boxed{\begin{split}
  I_{1,3} = \frac{a^{n-1} b^{1-2 n} \Gamma (1-n) \Gamma \left(n-\frac{1}{2}\right)\, _{1}F_{0}\left(n-\frac{1}{2}; ;\frac{a c}{b^2}\right)}{2 \sqrt{\pi }}+
    \frac{c^{1-n} \Gamma (n-1) \, _2F_1\left(\frac{1}{2},1;2-n;\frac{a
   c}{b^2}\right)}{2 b \Gamma (n)}, \\\\ \Big(a\neq0,c\neq0, b\neq0, \Big|\frac{ac}{b^{2}}\Big|< 1 \Big).\\\\
   \end{split}}
\end{equation}
The above solution is  valid for n \(\notin\) \( \mathbb{N}\) as there are Gamma functions which blow up for any \(n\in \mathbb{N}\).
Using eq. \eqref{entry1ans} and \eqref{entry1n2} we can write the follwing identity
\begin{equation}\label{newrelatn1}
    \boxed{\begin{split}
    \frac{(-1)^{n-1}}{•(n-1)!}\frac{\partial^{n-1}}{\partial c^{n-1}}\Big[\frac{1}{\sqrt{ac-b^2}}\,\cot^{-1}\frac{b}{\sqrt{ac-b^2}}\Big]= \frac{\sqrt{\pi } c^{\frac{1}{2}-n} \Gamma \left(n-\frac{1}{2}\right)\, _{1}F_{0}\left(n-\frac{1}{2}; ;\frac{b^2}{a c}\right)}{2 \sqrt{a} \Gamma (n)}\\\\
    -\frac{b c^{-n} \, _2F_1\left(1,n;\frac{3}{2};\frac{b^2}{a c}\right)}{a}.  
    \end{split}}
\end{equation}
The above equation relates the \(n^{th}\) derivative of the function appearing on LHS to the hypergeometric function \(_2F_1\), giving us a new identity.

\begin{center}
 \Large{2. Entry 3.252-3  } 
\end{center}
\vspace{.5cm}
We consider the following integral
    \begin{equation} \label{entry3}
        I=\int_{0}^{\infty}\frac{dx}{(ax^2+2bx+c)^{n+\frac{3}{2}}}
    \end{equation}
The value of this integral, as has been mentioned in Gradshteyn and Ryzhik is
    \begin{equation} \label{entry3ans}
        \int_{0}^{\infty}\frac{dx}{(ax^2+2bx+c)^{n+\frac{3}{2}}}=\frac{(-2)^n}{(2n+1){!}{!}}\frac{\partial^n}{\partial c^n}\Big(\frac{1}{\sqrt{c}(\sqrt{ac}+b)}\Big),\quad [a\geq0,\quad c>0,\quad b>-\sqrt{ac}].
    \end{equation}
 We expand the denominator as a bracket series:
    \begin{equation} 
        (ax^2+2bx+c)^{-n-\frac{3}{2}}=\sum_{n_1,n_2,n_3}^{\infty}\phi_{1,2,3} a^{n_1}(2b)^{n_2}c^{n_3}x^{2n_1+n_2}\frac{\langle n+\frac{3}{2}+n_1+n_2+n_3\rangle}{\Gamma(n+\frac{3}{2}).}
    \end{equation}
    Substituting the above expansion in eq. \eqref{entry3}, and integrating over \(x\) we obtain following bracket series:
    \begin{equation} \label{brac3}
        I=\sum_{n_1,n_2,n_3}^{\infty}\phi_{1,2,3} a^{n_1}(2b)^{n_2}c^{n_3}\frac{\langle 2n_1+n_2+1\rangle \langle n+\frac{3}{2}+n_1+n_2+n_3 \rangle}{\Gamma(n+\frac{3}{2}).}
    \end{equation}
    Following is the system of linear equations to be solved:
    \begin{align*} 
    2n_1 + n_2 + 1 &= 0, \\  
    n + \frac{3}{2} + n_1  + n_2 + n_3 &= 0.
    \end{align*}
There are 3 solutions taking one free variable each time.  Following are the solutions obtained.\\\\
      \textbf{1) \(n_2\) as the free variable\\\\}
        \begin{equation}\label{entry3_n2}
       \boxed{  \begin{split}
           I_2= \frac{c^{-n}}{{2 a \left(a c-b^2\right)}} \left(\frac{\sqrt{\pi } a^{3/2} \Gamma (n+1) \,_1F_0\left(n; ;\frac{b^2}{ac}\right)}{\Gamma \left(n+\frac{3}{2}\right)}+\frac{2 \left(b^3-a b c\right) \,
   _2F_1\left(1,n+\frac{3}{2};\frac{3}{2};\frac{b^2}{a c}\right)}{c^{3/2}}\right),\\\\
   \Big(c>0, \Big|\frac{b^2}{ac}\Big|< 1\Big).
   \end{split}}
        \end{equation}
Above solution is for all \(n \in \mathbb{R}
^+\) \\\\
\noindent
\textbf{2) \(n_1\) and \(n_3\) as the free variable\\\\}
The solution obtained using \(n_1\) and \(n_3\) have the same region of convergence hence they both are to be added to get the full answer. The final solution is:
        \begin{equation}\label{entry3n1_n3}
       \boxed{ \begin{split}
        I_{1,3}= \frac{1}{2} \Gamma \left(-n-\frac{1}{2}\right) \left(\frac{a^{n+\frac{1}{2}} b^{-2 n} \Gamma (n+1) _{1}F_{0}\left(n; ;\frac{a c}{b^2}\right)}{\sqrt{\pi }
   \left(b^2-a c\right)}-\frac{c^{-n-\frac{1}{2}} \, _2\tilde{F}_1\left(\frac{1}{2},1;\frac{1}{2}-n;\frac{a c}{b^2}\right)}{b}\right),\\\\
   \Big(b>0, \big|\frac{ac}{b^2}\big|< 1\Big).
   \end{split}}
    \end{equation}
    where, \(_2\tilde{F}_1\) is the regularized Hypergeometric \(_2F_1\), defined as follows:
    \begin{equation*}
        \, _2\tilde{F}_1(a,b;c;z)=\frac{\, _2F_1(a,b;c;z)}{\Gamma (c)}
    \end{equation*}
    above solution is for all \(n\in \mathbb{R}^+\) except half-integers \(\frac{1}{2},\frac{3}{2},...etc\)\\\\
Using eq. \eqref{entry3ans} and \eqref{entry3_n2} we can also write the following relation\\
\begin{equation}\label{newrelatn2}
   \boxed{ \begin{split}
    \frac{(-2)^n}{(2n+1){!}{!}}\frac{\partial^n}{\partial c^n}\Big(\frac{1}{\sqrt{c}(\sqrt{ac}+b)}\Big)=  \frac{c^{-n}}{{2 a \left(a c-b^2\right)}} \Bigg(\frac{\sqrt{\pi } a^{3/2} \Gamma (n+1) \,_1F_0\left(n; ;\frac{b^2}{ac}\right)}{\Gamma \left(n+\frac{3}{2}\right)}\\
    \hspace{2cm}+\frac{2 \left(b^3-a b c\right) \,
   _2F_1\left(1,n+\frac{3}{2};\frac{3}{2};\frac{b^2}{a c}\right)}{c^{3/2}}\Bigg)
\end{split}}
\end{equation}
Above equation gives us a new identity for the hypergeometric function, relating the \(n^{th}\) derivative of the function appearing on LHS to \(_2F_1\) hypergeometric function.
\newpage

\begin{center}
   \Large{3. Entry 3.252-4}
\end{center}
 The integral to be considered is
    \begin{equation} \label{entry4}
        I= \int_{0}^{\infty}\frac{x\, dx}{(ax^2+2bx+c)^{n}}
    \end{equation}
    \\
    The value of this integral, as has been mentioned in Gradshteyn and Ryzhik is
\begin{equation} \label{entry4ans}
\begin{split}
        \int_{0}^{\infty}dx\,\frac{x}{(ax^2+2bx+c)^{n}} & = \frac{(-1)^n}{(n-1){!}{!}}\frac{\partial^{n-2}}{\partial c^{n-2}}\Big(\frac{1}{2(ac-b^2)}-\frac{b}{2(ac-b^2)^{\frac{3}{2}}} \cot^{-1}(\frac{b}{\sqrt{ac-b^2}})\Big),[ac>b^2], \\
       & = \frac{(-1)^n}{(n-1){!}{!}}\frac{\partial^{n-2}}{\partial c^{n-2}}\Big(\frac{1}{2(ac-b^2)}+\frac{b}{4(b^2-ac)^{\frac{3}{2}}} \ln\Big(\frac{b+\sqrt{b^2-ac}}{b-\sqrt{b^2-ac}}\Big)\Big),
       [b^2>ac>0], \\
 &= \frac{a^{n-2}}{2(n-1)(2n-1)b^{2n-2}},\quad[ac=b^2].
\end{split}
\end{equation}
Expanding the denominator as a bracket-series:
\begin{equation}
    (ax^2+2bx+c)^{-n}=\sum_{n_1,n_2,n_3}^{\infty}\phi_{1,2,3} a^{n_1}(2b)^{n_2}c^{n_3}x^{2n_1+n_2}\frac{\langle n+n_1+n_2+n_3\rangle}{\Gamma(n)}.
\end{equation}
 Substituting the above expansion in eq. \eqref{entry4} and integrating over \(x\) we get the following bracket series
    \begin{equation}\label{brac4}
        I=\sum_{n_1,n_2,n_3}^{\infty}\phi_{1,2,3}a^{n_1}(2b)^{n_2}c^{n_3}\frac{\langle 2n_1+n_2+2\rangle \langle n+n_1+n_2+n_3\rangle}{\Gamma(n)}.
    \end{equation}
    Following is the system of linear equations to be solved
\begin{align*} 
    2n_1 + n_2 + 2 &= 0, \\  
    n + n_1  + n_2 + n_3 &= 0.
\end{align*}
   There are three different solutions taking one variable to be free at a time.
\noindent
Following are the solutions obtained:\\\\
\textbf{1) \(n_2\) as the free variable\\}
\begin{equation}\label{entry_n2}
\boxed{ \begin{split}
I_2=\frac{c^{1-n} \Gamma (n-1) \, _2F_1\left(1,n-1;\frac{1}{2};\frac{b^2}{a c}\right)}{2 a \Gamma (n)}-\frac{\sqrt{\pi } b c^{\frac{1}{2}-n} \Gamma \left(n-\frac{1}{2}\right)\, _{1}F_{0}\left(n-\frac{1}{2}; ;\frac{b^2}{ac}\right)}{2 a^{3/2} \Gamma (n)},\\
\hspace{.6cm}\Big(c>0,b>0,\Big|\frac{b^2}{ac}\Big|<1\Big).\\\\
 \end{split}}
 \end{equation}
above solution is for all  \(n \in\) \( \mathbb{R}^{+}\) \\\\
 \textbf{2) \(n_1\) and \(n_3\) as the free variable \\}
        \ \begin{equation}\label{entryn1_n3}
        \boxed{\begin{split}
            I_{1,3}=-\frac{\Gamma (1-n) \left(2 a^{n-2} b^{4-2 n} \Gamma \left(n-\frac{1}{2}\right) \,_{1}F_{0}\left(n-\frac{1}{2}; ;\frac{a c}{b^2}\right)-\sqrt{\pi } c^{2-n}
   \, _2\tilde{F}_1\left(1,\frac{3}{2};3-n;\frac{a c}{b^2}\right)\right)}{4 \sqrt{\pi } b^2},\\\\
   \Big(a>0, b> 0 , \Big|\frac{ac}{b^2}\Big|<1 \Big).
   \end{split}}
   \end{equation}
 above solution is for all n \(\notin\) \( \mathbb{N}\)\\
   For the case where \(ac = b^2\) we substitute the above special condition in equation eq. \eqref{entry4} and proceed as before.\\
\noindent
Using eq. \eqref{entry4ans} and \eqref{entryn1_n3} we can write the following identity\\
\begin{equation}\label{newrelatn3}
   \boxed{ 
   \begin{split}
    \frac{(-1)^n}{(n-1){!}{!}}\frac{\partial^{n-2}}{\partial c^{n-2}}\Big(\frac{1}{2(ac-b^2)}-\frac{b}{2(ac-b^2)^{\frac{3}{2}}} \cot^{-1}(\frac{b}{\sqrt{ac-b^2}})\Big)= \frac{c^{1-n} \Gamma (n-1) \, _2F_1\left(1,n-1;\frac{1}{2};\frac{b^2}{a c}\right)}{2 a \Gamma (n)}\\
    -\frac{\sqrt{\pi } b c^{\frac{1}{2}-n} \Gamma \left(n-\frac{1}{2}\right)\, _{1}F_{0}\left(n-\frac{1}{2}; ;\frac{b^2}{ac}\right)}{2 a^{3/2} \Gamma (n)}.   
    \end{split}
    }
\end{equation}
Above equation is a new identity for the hypergeometric function \(_2F_1\).

\subsection{Generalization}
We now present a generalization of the quadratic type integrals with some general powers of numerator and denominator given by
\begin{equation}\label{genquad}
I = \int_{0}^{\infty}\frac{x^{n}\,dx}{(ax^2 + 2bx +c)^{m}}. 
\end{equation}
First step is to expand the denominator as a bracket series.
\begin{equation*}
(ax^2 + 2bx +c)^{-m} = \sum_{n_{1},n_{2},n_{3}}\phi_{1,2,3}\frac{(ax^2)^{n_{1}}(2bx)^{n_{2}}(c)^{n_{3}}\langle m+n_{1}+n_{2}+n_{3}\rangle}{\Gamma(m)}.
\end{equation*}
Substituting the above expansion in the eq. \eqref{genquad} and integrating over \(x\) we get the following bracket series.
\begin{equation*}
I = \sum_{n_{1},n_{2},n_{3}}\phi_{1,2,3}\frac{(a)^{n_{1}}(2b)^{n_{2}}(c)^{n_{3}}\langle m+n_{1}+n_{2}+n_{3}\rangle\langle{}2n_{1}+n_{2}+n+1\rangle}{\Gamma(m)}.
\end{equation*}
We have the following two linear equations to solve
\begin{align*} 
m + n_1 + n_2 +n_3 = 0,\\
2n_1 + n_2 + n + 1 = 0.
\end{align*}
There are three variables and two equation so there are 3 different solution by taking one free variable each time. Following are the solutions obtained\\\\
\textbf{1) \(n_2\) as the free variable}
\begin{equation}\label{genquadn22}
\boxed{\begin{split}
I_2= \frac{a^{-\frac{n}{2}-\frac{1}{2}} \Gamma \left(\frac{n+1}{2}\right) c^{\frac{1}{2} (n-2 m)+\frac{1}{2}} \Gamma \left(m-\frac{n}{2}-\frac{1}{2}\right) \,
   _2F_1\left(m-\frac{n}{2}-\frac{1}{2},\frac{n+1}{2};\frac{1}{2};\frac{b^2}{a c}\right)}{2 \Gamma (m)}\\\\
   -\frac{b a^{-\frac{n}{2}-1} \Gamma \left(\frac{n}{2}+1\right) c^{\frac{1}{2} (n-2 m)} \Gamma
   \left(m-\frac{n}{2}\right) \, _2F_1\left(m-\frac{n}{2},\frac{n}{2}+1;\frac{3}{2};\frac{b^2}{a c}\right)}{\Gamma (m)},\\
   \Big(\Big|\frac{b^2}{ac}\Big|< 1\Big).
 \end{split}}
\end{equation}
\textbf{2) \(n_1\) and \(n_3\) as the free variables}\\\\
The solution corresponding to \(n_1\) and \(n_3\) as the  free variables have the same region of convergence and are added together to get the full answer.  The final answer after simplification is\\
\begin{equation}\label{genquadn1_n3}
\boxed{\begin{split}
   I_{1,3}= \frac{2^{-2 m+n+1} a^{m-n-1} b^{-2 m+n+1} \Gamma (2 m-n-1) \Gamma (-m+n+1) \, _2F_1\left(m-\frac{n}{2}-\frac{1}{2},m-\frac{n}{2};m-n;\frac{a c}{b^2}\right)}{\Gamma (m)}\\\\
   +\frac{2^{-n-1} b^{-n-1} \Gamma
   (n+1) c^{-m+n+1} \Gamma (m-n-1) \, _2F_1\left(\frac{n+1}{2},\frac{n+2}{2};-m+n+2;\frac{a c}{b^2}\right)}{\Gamma (m)},\\\\
   \Big(\Big|\frac{ac}{b^2}\Big|<1\Big).
\end{split}}
\end{equation}
The above evaluation for some general powers \(n\) and \(m\) is not given in Gradshteyn and Ryzhik. Putting special values for\(m\) and \(n\) will give results for entries- 1, 4, 7, 8 and 9 as the special case.

\subsection{Special case n= 0}
\textbf{1) \(n_2\) as the free variables\\\\}
Simplifying we get 
\begin{equation}
\begin{split}
I_2= \frac{\sqrt{\pi } c^{\frac{1}{2}-m} \Gamma \left(m-\frac{1}{2}\right) \,_{1}F_{0}\left(m-\frac{1}{2}; ;\frac{b^2}{ac}\right)}{2 \sqrt{a} \Gamma (m)}-\frac{b c^{-m} \, _2F_1\left(1,m;\frac{3}{2};\frac{b^2}{a
   c}\right)}{a}.\\\\
\end{split}
\end{equation}
which is same as eq. \eqref{entry1n2}\\\\
\textbf{2) \(n_1\) and \(n_3\) as the free variables}\\\\
Simplifying we get
\begin{equation}
\begin{split}
  I_{1,3}= \frac{2^{1-2 m} a^{m-1} b^{1-2 m} \Gamma (1-m) \Gamma (2 m-1) _{1}F_{0}\left(m-\frac{1}{2}; ;\frac{a c}{b^2}\right)+\frac{c^{1-m} \Gamma (m-1) \, _2F_1\left(\frac{1}{2},1;2-m;\frac{a c}{b^2}\right)}{2
   b}}{\Gamma (m)}.
   \end{split}
\end{equation}\\
using the Legendre duplication formula on \(\Gamma(2m-1)\) we get eq. \eqref{entry1n3&n4}.  Similarly putting different values of \(n\) and \(m\), we can get the other entries.

\section{Quartic integral}

The quartic integral is given by 
\begin{equation}\label{quartic}
    I = \int_{0}^{\infty}\frac{dx}{(a x^4 + 2bx^2 +c)^m}.
\end{equation}
The only appearance of the above integral in the table is as entry 2.161-5 in terms of recursive integrals. The above integral has been evaluated for the special case of \(a= 1\) and \(c= 1\) in \cite{Gonzalez:part1} using the method of brackets. Here we evaluate it for some general a and c.
First we expand the denominator as a bracket series
\begin{equation*}
(ax^4 + 2bx^2 +c)^{-m} = \sum_{n_{1},n_{2},n_{3}}\phi_{1,2,3}\frac{(ax^4)^{n_{1}}(2bx^{2})^{n_{2}}(c)^{n_{3}}\langle m+n_{1}+n_{2}+n_{3}\rangle}{\Gamma(m)}.
\end{equation*}
Substituting the above expansion in the eq. \eqref{quartic} and integrating over \(x\) we get the following bracket series.
\begin{equation*}
I = \sum_{n_{1},n_{2},n_{3}}\phi_{1,2,3}\frac{(a)^{n_{1}}(2b)^{n_{2}}(c)^{n_{3}}\langle m+n_{1}+n_{2}+n_{3}\rangle\langle4n_{1}+2n_{2}+1\rangle}{\Gamma(m)}.
\end{equation*}
We have the following two linear equations to solve
\begin{align*} 
m + n_1 + n_2 +n_3 = 0,\\
4n_1 + 2n_2 + 1 = 0.
\end{align*}
There are three variables and two equation so there are 3 different solution by taking one free variable each time. Following are the solutions obtained\\\\
\textbf{1) \(n_2\) as the free variable}
\begin{equation}\label{quartic_n2}
\boxed{\begin{split}
I_2= \frac{\Gamma \left(\frac{1}{4}\right) c^{\frac{1}{4}-m} \Gamma \left(\frac{1}{4} (4 m-1)\right) \, _2F_1\left(\frac{1}{4},m-\frac{1}{4};\frac{1}{2};\frac{b^2}{a c}\right)}{4 \sqrt[4]{a} \Gamma
   (m)}-\frac{b \Gamma \left(\frac{3}{4}\right) c^{-m-\frac{1}{4}} \Gamma \left(\frac{1}{4} (4 m+1)\right) \, _2F_1\left(\frac{3}{4},m+\frac{1}{4};\frac{3}{2};\frac{b^2}{a c}\right)}{2 a^{3/4} \Gamma
   (m)},\\\\
   \Big(\Big|\frac{b^2}{ac}\Big|< 1\Big).
 \end{split}}
\end{equation}
\textbf{2) \(n_1\) and \(n_3\) as the free variables}\\\\
The solution corresponding to \(n_1\) and \(n_3\) as the  free variables have the same region of convergence  and are added together to get the full answer.  The final answer after simplification is\\\\
\begin{equation}\label{quarticn1_n3}
\boxed{\begin{split}
   I_{1,3}= \frac{2^{-2 m-\frac{1}{2}} a^{m-\frac{1}{2}} b^{\frac{1}{2}-2 m} \Gamma \left(\frac{1}{2} (1-2 m)\right) \Gamma \left(\frac{1}{2} (4 m-1)\right) \,
   _2F_1\left(m-\frac{1}{4},m+\frac{1}{4};m+\frac{1}{2};\frac{a c}{b^2}\right)}{\Gamma (m)}\\\\
   +\frac{\sqrt{\frac{\pi }{2}} c^{\frac{1}{2}-m} \Gamma \left(\frac{1}{2} (2 m-1)\right) \,
   _2F_1\left(\frac{1}{4},\frac{3}{4};\frac{3}{2}-m;\frac{a c}{b^2}\right)}{2 \sqrt{b} \Gamma (m)},\\\\
   \Big(\Big|\frac{ac}{b^2}\Big|<1\Big).
\end{split}}
\end{equation}
With \(a=c=1\) we will reproduce the result evaluated using the method of brackets after some simplification.

\subsection{Generalization}
Next we evaluate a more general quartic integral given by
\begin{equation}\label{genquartic}
    I = \int_{0}^{\infty}\frac{dx}{x^n(a x^4 + 2bx^2 +c)^m}.
\end{equation}
The above integral occurs as a recursion relation in entry 2.161-6 of the table.
First we expand the denominator as a bracket series
\begin{equation*}
(ax^4 + 2bx^2 +c)^{-m} = \sum_{n_{1},n_{2},n_{3}}\phi_{1,2,3}\frac{(ax^4)^{n_{1}}(2bx^{2})^{n_{2}}(c)^{n_{3}}\langle m+n_{1}+n_{2}+n_{3}\rangle}{\Gamma(m)}.
\end{equation*}
Substituting the above expansion in the eq. \eqref{genquartic} and integrating over \(x\) we get the following bracket series.
\begin{equation*}
I = \sum_{n_{1},n_{2},n_{3}}\phi_{1,2,3}\frac{(a)^{n_{1}}(2b)^{n_{2}}(c)^{n_{3}}\langle m+n_{1}+n_{2}+n_{3}\rangle\langle4n_{1}+2n_{2}-n+1\rangle}{\Gamma(m)}.
\end{equation*}
We have the following two linear equations to solve
\begin{align*} 
m + n_1 + n_2 +n_3 = 0,\\
4n_1 + 2n_2 - n + 1 = 0.
\end{align*}
There are three variables and two equation so there are 3 different solution by taking one free variable each time. Following are the solutions obtained\\\\
\textbf{1) \(n_2\) as the free variable}
\begin{equation}\label{genquartic_n2}
\boxed{
\begin{split}
I_2= a^{\frac{n-3}{4}} c^{\frac{1}{4} (-4 m-n-1)} \Bigg(\frac{\sqrt{a} \sqrt{c} \Gamma \left(\frac{1}{4}-\frac{n}{4}\right) \Gamma \left(m+\frac{n}{4}-\frac{1}{4}\right) \,
   _2F_1\left(\frac{1-n}{4},\frac{1}{4} (4 m+n-1);\frac{1}{2};\frac{b^2}{a c}\right)}{4 \Gamma (m)}\\
   -\frac{b\, \Gamma \left(\frac{3}{4}-\frac{n}{4}\right) \Gamma \left(m+\frac{n}{4}+\frac{1}{4}\right) \,
   _2F_1\left(\frac{3}{4}-\frac{n}{4},m+\frac{n}{4}+\frac{1}{4};\frac{3}{2};\frac{b^2}{a c}\right)}{2 \Gamma (m)}\Bigg),\\
   \Big(\Big|\frac{b^2}{ac}\Big|< 1\Big).
 \end{split}}
\end{equation}
\textbf{2) \(n_1\) and \(n_3\) as the free variables}\\\\
The solution corresponding to \(n_1\) and \(n_3\) as the  free variables have the same region of convergence  and are added together to get the full answer.  The final solution after simplification is\\\\
\begin{equation}\label{genquarticn1_n3}
\boxed{\begin{split}
   I_{1,3}= 2^{-\frac{n}{2}-\frac{3}{2}} b^{-\frac{n}{2}-\frac{1}{2}} \Bigg(\frac{2^{1-2 m} b^{1-2 m} a^{\frac{1}{2} (2 m+n-1)} \Gamma \left(\frac{1}{2} (-2 m-n+1)\right) \Gamma \left(\frac{1}{2} (4 m+n-1)\right) \,
  }{\Gamma (m)}\\
  \times\, _2F_1\left(m+\frac{n}{4}-\frac{1}{4},m+\frac{n}{4}+\frac{1}{4};m+\frac{n}{2}+\frac{1}{2};\frac{a c}{b^2}\right)\\\\
   +\frac{2^n b^n \Gamma \left(\frac{1-n}{2}\right) c^{\frac{1}{2} (-2 m-n+1)}
   \Gamma \left(\frac{1}{2} (2 m+n-1)\right) \, _2F_1\left(\frac{1}{4}-\frac{n}{4},\frac{3}{4}-\frac{n}{4};-m-\frac{n}{2}+\frac{3}{2};\frac{a c}{b^2}\right)}{\Gamma (m)}\Bigg) \Bigg),\\\\
   \Big(\Big|\frac{ac}{b^2}\Big|<1\Big).
\end{split}}
\end{equation}
Putting \(n=0\) in eq. \eqref{genquarticn1_n3} and \eqref{genquartic_n2} we get the previous results.
\section{Conclusion}
The method of brackets has some popularity in the evaluation of Feynman integrals
having its origins in the negative dimension integration method, and rests on the
Ramanujan's master theorem.  It has also been applied in the past to definite
integrals on the positive real line from time to time.  These integrals have a life
of their own as well.  In the present work, motivated by several considerations,
we have solved for definite integrals of the quadratic and quartic type that
have made an appearance in the tables by Gradshyteyn and Ryzhik.\newline \indent
The quadratic integrals are given in terms of \(n^{th}\) partial derivative of a function in the table by Gradshteyn and Ryzhik.  We were able to evaluate all those integrals and obtain closed form expression in terms of hypergeometric function.  We have considered further
generalizations of these, and have provided expressions  in terms
of hypergeometric functions resulting from the use of the method.  These generalizations allow us to evaluate various entries as the special case with some special values of the parameters \(a, b, c\) and \(n\).  We have also considered quartic type integrals for general values of parameters \(a, b, c\) and \(n\) and then considered a further generalization of this integral.
For quadratic type integrals the new results supersede
the results in some cases of Gradshteyn and Ryzhik which are given only for natural numbers, while
our results hold for any positive real.  By equating the new expressions with the known
results, we therefore present new inter-relations between these specific
hypergeometric functions appearing in our results. For the quartic type integrals the results in the table by Gradshteyn and Ryzhik are given only as the recursive relation.  The new relation obtained using the method of brackets can also allow us to relate the two.  We have also provided some
simple examples as an illustration of the method for purposes of pedagogy.
Thus our work is a contribution in the line of recent results that have appeared
in the literature. \\\\
\section{Acknowledgement}
We thank Samuel Friot for several interesting discussions on
the subject.  We also thank Victor H. Moll for his constant encouragement and
for replying to all our queries.

\end{document}